\documentclass[aps,prx,reprint,superscriptaddress,longbibliography,amsmath,amssymb]{revtex4-1}
\usepackage[utf8]{inputenc}
\usepackage{amsfonts}
\usepackage{graphicx} 
\usepackage{color}
\usepackage[dvipsnames]{xcolor}
\usepackage{hyperref}
\usepackage{comment}

\newcommand{\Neel}{N{\'e}el}
\newcommand{\dJ}[1]{{\color{blue}$\clubsuit$ dj: #1}}

\usepackage{soul}

\newcommand{\csm}{Department of Physics, Colorado School of Mines, Golden, Colorado 80401, USA}
\newcommand{\luh}{Institut f{\"u}r Theoretische Physik Leibniz Universi{\"a}t Hannover, Appelstrasse 2, 30167 Hannover, Germany}

\newcommand{\HtoN}{(A)} 
\newcommand{\HtonH}{(B)} 

\begin{document}

\title{Dynamics for the Haldane phase in the Bilinear-Biquadratic Model}

\author{Arya Dhar}
\affiliation{\csm}
\affiliation{\luh}
\author{Daniel Jaschke}
\affiliation{\csm}
\author{Lincoln D. Carr}
\affiliation{\csm}


\begin{abstract}
  The BBM is a promising candidate to study spin-one
  systems and to design quantum simulators based on its underlying Hamiltonian. The
  variety of different phases contains amongst other valuable and exotic phases the
  Haldane phase. We study the Kibble-Zurek physics of linear quenches into the Haldane phase. We outline ideal quench protocols to minimize
  defects in the final state while exploiting different linear quench
  protocols via the uniaxial or interaction term. Furthermore, we look at the fate of the string order
  when quenching from a topologically non-trivial phase to a trivial phase. Our studies show this depends significantly on the path chosen for quenching;
  for example, we discover quenches from \Neel{} to Haldane phase which reach a
  string order greater than their ground state counterparts for the initial
  or final state at intermediate quench times.
\end{abstract}

\maketitle

\section{Introduction                                                           \label{sec:intro}}

The last three decades have witnessed an unprecedented progress~\cite{altman2019quantum} in fulfilling Feynman's vision 
of constructing a quantum simulator which would be able to solve quantum mechanical problems directly~\cite{Feynman1982}. 
One popular approach to achieve this goal is adiabatic quantum computing, which relies 
on the preparation of the ground state of a Hamiltonian that is easy to attain experimentally.
The system is then evolved adiabatically to the final Hamiltonian whose ground state
encodes the solution to a particular computational or optimization problem~\cite{Lidar_RMP2018}. 
Adiabatic quantum computing has been used in the D-wave architecture and is strongly related
to quenches through a quantum critical point and spin-$1/2$ models. However, we are not constrained to spin-$1/2$ systems: spin-1 and beyond have even more intriguing features.

Quantum phases are usually characterized by local order parameters, and phase transitions
are then described by symmetry breaking according to Landau's theory. However, there is a different class
of phases called \emph{topological phases} which are characterized by non-local order parameters. These
phases often have a gap in the bulk energy spectrum with gapless modes residing at the edges.
Furthermore, a new class of topological phase was discovered in the last
decade possessing symmetry-protected topological order in which the
gapless edge excitations are preserved by
symmetries~\cite{Wen_PRB2009, Senthil_ACP2015}. A paradigmatic example
of a symmetry-protected topological (SPT) phase is the Haldane phase
exhibited by the Heisenberg model with odd integer
spins~\cite{HaldanePLA1983}. This phase is protected by three symmetries:
time reversal, spatial inversion and $Z_2 \times Z_2$ symmetry. The phase
possesses a non-local string order.  These SPT phases have become potential
candidates for measurement-based quantum computation for which SPT
order ensures the perfect operation of the identity
gate~\cite{Doherty_PRL2012, Raussendorf_PRL2017}. There have also been
proposals to use SPT phases as adiabatic quantum transistors which are
universal adiabatic quantum computing devices whose operational speed
depends on the minimal energy gap~\cite{Crosswhite_PRX2015, Bartlett_NJP2015}.
These proposed logic gates, due to their symmetry-protected feature, have
been argued to be quite robust against a variety of relevant noise processes.
Recently, a metrological application with the Haldane insulator was
proposed in which the passive, error-preventing properties of the SPT
phase can be used to measure the direction of an unknown electric
field~\cite{Bartlett_QST2017}.

Ultracold gases loaded in optical lattices offer an exceptionally
high degree of controllability over the geometry and interactions
as well as time-dependent quenches. Recent years have seen
a remarkable development in the variety of non-equilibrium
experiments achieved by the ultracold gases such as studying
their transport properties~\cite{SchneiderPRL2013,AidelsburgerPRL2018},
thermalization~\cite{KinoshitaNature2006}, many-body
localization~\cite{BlochScience2015, ScheiderPRX2017}, relaxation
dynamics~\cite{BeugnonPRL2017}, and quench dynamics across a phase
transition~\cite{LamporesiNatPhys2013, MeldginNatPhys2016}. The
recent realization of
quantum integer-spin chains with tunable interactions using trapped
ions opens up possibilities to study SPT phases in
spin-1 systems~\cite{Monroe_PRX2015}. Ultracold molecules represent
an alternative way to achieve these effective three level
systems, also called qutrits, via the large number of hyperfine
levels and the electric dipole moment~\cite{Carr_NJP2009, Jin_CC2017}. External
magnetic and laser fields can control the various
interaction terms appearing in the desired Hamiltonian and perform
slow to fast quenches. The development of the quantum gas
microscope with single site and spin resolution will grant access to
the measurement of the local and non-local order
parameters~\cite{Bakr_Nature2009, Bloch_Nature2010, Endres2011}.

The rapid development in the field of quantum computation and its
relation to the SPT phases, such as the Haldane phase and the
subsequent progress in experimental endeavours have prompted us to
analyze in detail quench dynamics across quantum critical points
in the spin-one bilinear-biquadratic model (BBM) associated with
the Haldane phase~\cite{Demler_PRA2003,Nomura_PRB2006, Trebst_PRB2006,
Vekua_PRL2011, DeChiara2011}.  Spin-one models,
i.e., qutrits, also enable more powerful applications due
to more internal degrees of freedom per site~\cite{low2019practical}. In spite of many
theoretical predictions, the Haldane phase has remained elusive in experiments. Through  careful analysis, we propose in this Article
the parameters for linear quench protocols which minimize defect generation and thus
will provide experimentalists a route to observe the Haldane
phase with a finite string order. The fate of string order when
quenching is examined through different pathways to a topologically non-trivial phase; 
up till now, the dynamics of the Haldane phase has still not been studied in much detail. Apart from the experimental
stimulus, there is also the motivation of relating defect
generation to the quench speed. For quenches across second-order
phase transitions, the Kibble-Zurek hypothesis proposes a universal
nature of the density of defects, relating them to the critical
exponents of the underlying quantum phase transition. We analyze the
statics and dynamics of the BBM with the
matrix product state (MPS) method \cite{Schollwoeck2011} which is
well-suited for 1-dimensional entangled many-body systems and
gives us access to a variety of relevant measures that can be
pivotal in the analysis.


The paper is arranged as follows: we begin with the definition of the BBM in
Sec.~\ref{sec:model}, also containing a general discussion of the Kibble-Zurek mechanism. The results of the different quench protocols to the Haldane phase are presented in Sec.~\ref{sec:results}.
We elaborate the methods, detailed error analysis and the static results for finite-size systems
in Sec.~\ref{sec:methods}.
Finally, Sec.~\ref{sec:concl} concludes our work with discussions.

\section{Model                                                                  \label{sec:model}}

Although our results are general to many quantum simulator architectures, we contextualize our study with the specific architecture of an ultracold spin-1 bosonic gas trapped in
a 1-dimensional optical lattice with repulsive interactions
between them. Because of the identity of bosons undergoing an $s$-wave
interaction, the total spin of the two interacting bosons, $S_{tot}$ can
be $0,2$. If the tunneling amplitude $t$ between neighboring lattice sites
is small and finite, one can apply a second order perturbation theory
in $t$ to get the low energy physics which is given by the superexchange
processes. The corresponding spin Hamiltonian for two neighboring sites at
unit filling can be expanded in powers of the nearest-neighbor Heisenberg
interactions, $H_{i,i+1}=\sum_k a_k(\textbf{S}_i.\textbf{S}_{i+1})^k$.
Terminating the series at $k=2$, i.e., second order perturbation theory, gives the BBM, up to the external field which is often  also included in the BBM. The
presence of a symmetry-breaking field can have important consequences.  The
linear Zeeman effect does not play a role since it can be gauged out due
to the fact that the total magnetization is a constant of motion. It should be noted that this symmetry is used as well for the numerical simulations. On the
other hand, the quadratic Zeeman effect leads to effects in spinor gases which cannot be gauged away. Combining the two effects leads
us to the BBM with quadratic Zeeman field:
\begin{eqnarray}                                                                \label{eq:HBBM}
  \mathcal{H} &=& J \sum_{i=1}^{L-1} [ \cos(\theta)\textbf{S}_i.\textbf{S}_{i+1}
    + \sin(\theta) (\textbf{S}_i.\textbf{S}_{i+1})^2] \nonumber \\
    &&+ D \sum_{i=1}^{L} {{}{\hat{S}}_{i}^{z}}^{2}, 
\end{eqnarray}
where $\textbf{S}_i=(\hat{S}^x_{i}, \hat{S}^y_{i}, \hat{S}^z_{i})$
are the angular momentum operators located at the $i^\mathrm{th}$ site of a 1D $L$-site qutrit, or spin-one lattice. The first sum in Eq.~(\ref{eq:HBBM}) is the bilinear-biquadratic part tuned by the parameter $\theta$,
whereas the second term is the uniaxial field $D$ taking into account
the quadratic Zeeman field. Some of the phases exhibited by Eq.~(\ref{eq:HBBM}) have degenerate ground states. Hence, for the purposes of numerical
calculations a very small symmetry-breaking field is applied to
the $L^\mathrm{th}$ site.

The Hamiltonian in Eq.~\eqref{eq:HBBM} obeys a $\mathcal{U}(1)$ symmetry
similar to the number conservation in the Bose-Hubbard model, where here
the total spin in the $z$-direction is conserved. The generator of the
symmetry is $\oplus_{i=1}^{L} \hat{S}^z_{i}$. Thus, the possible symmetry sectors
reach from a total $z$-spin of $-L$ to $+L$. We use a total spin of zero
throughout our simulations with open boundary conditions. We follow the
convention that the time is in units of the interaction $J$.

The BBM has been extensively studied revealing a plethora
of phases such as the dimer phase, the \Neel{} phase, the Haldane phase, to
name a few~\cite{DeChiara2011}. Due to the high relevance in recent
experiments with ultracold gases in optical lattices~\cite{Endres2011, Hilker2017},
we study the effects of quenches across
the following phase boundaries shared by the Haldane phase:
\begin{itemize}
  \item \HtoN{} - \Neel{} and Haldane phase
  \item \HtonH{} - Large-D (non-Haldane) and Haldane phase
\end{itemize}
The critical points for the respective quantum phase transitions
for the specific lengths of the systems are first determined using the respective order parameters. The linear quenches start and
finish almost equidistant from the corresponding critical points on either side. We ensure that
the initial and final points are deep inside the respective phases, so that the effects
of the critical region are properly considered.

The effects of the linear quench, crossing the quantum critical point
in the process, are examined via the following six observables:

\begin{enumerate}
\item{The \emph{instantaneous energy gap} is a static measure characterizing the difference
  between the ground state energy $E_{0}$ of the system and the energy  $E_k$
  of the $k^\mathrm{th}$ relevant excited state compatible with the integrals of motion
  \begin{eqnarray}
    \Delta_{0k} = E_{k} - E_{0} \, .
    \label{eq:DynamicalGap}
  \end{eqnarray}
  The choice of the ground and relevant excited state considered for this
  measure always belong to the same symmetry sector, i.e., have the
  same spin projection in the $z$-direction.
}

\item{The \emph{staggered magnetization} is defined as
\begin{eqnarray}                                                                \label{eq:stagmag}
  M_z^{\mathrm{st}} = \frac{1}{L} \sum_{i=1}^{L} {(-1)^{i} \hat{S}^z_{i}} \, .
  \end{eqnarray}
  The \Neel{} phase is characterized by a finite value of the staggered magnetization.
  }

  \item{The \emph{entanglement entropy}, also called \emph{block entropy} is a
  measure defined based on the reduced density matrix or Schmidt
  decomposition, i.e.,
  \begin{eqnarray}                                                              \label{eq:BlockEntropy}
    S = -\mathrm{Tr}_{B} \rho_{AB} \log \rho_{AB}
      = -\mathrm{Tr} \rho_{A} \log \rho_{A} \, ,
  \end{eqnarray}
  where $\rho_A$ is the reduced density matrix of the subsystem $A$ by tracing
  over the degrees of freedom of the rest of the system, i.e., subsystem $B$.
  Throughout our work,
  we choose subsystem $A$ as the sites $1, \ldots, L / 2$ with $L$ being even; the sites
  $L / 2 + 1, \ldots L$ represent the subsystem $B$ with the degrees of freedom
  being traced out. Typically, the entanglement entropy scales with the size of the
  subsystem, $n_A \equiv L / 2$~\cite{KitaevPRL2003}. However, in one dimensional systems
  with only short-ranged terms in the Hamiltonian, the entanglement entropy
  saturates to a constant value independent of the size of the block. This
  property known as the ``area law'' \cite{Eisert2010}, is the cause behind the
  remarkable success of tensor-network-based methods in describing 1-dimensional systems. In the vicinity of a quantum critical point, the
  entanglement entropy starts to diverge logarithmically with the block size
  $n_{A}$ for large system sizes, as
  \begin{eqnarray}
    S \sim c \log (n_A) \, ,
  \end{eqnarray}
  where $c$ is the central charge of the conformal field theory describing
  the critical point~\cite{HolzheyNPB1994, KorepinPRL2004, CalabreseJSM2004,
  CalabreseJPA2009}. For systems slightly
  away from the critical point, when the correlation length $\xi$ in the
  ground state is large but finite, the entanglement entropy behaves as
  \begin{eqnarray}
    S \sim \mathcal{A}\frac{c}{6}\log \xi \, ,
    \label{eq:EntropyCorrlen}
  \end{eqnarray}
  where $\mathcal{A}$ is the number of boundary points of the
  system. 
}

\item{We consider the reduced density matrix of a subsystem or alternatively,
  the eigenvalues of the bipartitions via the singular values of the Schmidt decomposition, where the \emph{Schmidt gap}
  is defined as
  \begin{eqnarray}                                                              \label{eq:SchmidtGap}
    \Delta \lambda = \lambda_1 - \lambda_2 \, ,
  \end{eqnarray}
  with $\lambda_1, \lambda_2$ the highest two eigenvalues of the reduced
  density matrix.
  Using finite-size scaling, the Schmidt gap can 
  signal a quantum phase transition. $\Delta \lambda$ scales with the critical exponents
  related to the conformal field theory describing the transition
  point~\cite{SanperaPRB2013}. References~\cite{SanperaPRL2012, SanperaPRB2013}
  have recently shown that the Schmidt gap is related to the correlation length
  of the system,  $\xi$, through a power-law, where the exponent is called the dynamical critical exponent of the transition. Studying the complete entanglement spectrum in a dynamical problem can be complicated. Since both entanglement entropy and the Schmidt gap are related to the entanglement spectrum and give equivalent insights, we concentrated on these two established quantities.

}

\item{The \emph{string order parameter} is a non-local measurement
  acting on multiple sites defined as
  \begin{eqnarray}                                                              \label{eq:StringOrder}
    O^S &=& \lim_{L \to \infty} O_{i}(r=L-2i) \nonumber \\
     &\equiv & \lim_{r\rightarrow\infty} \left\langle
          \hat{S}^z_{i} \exp\left[ \mathrm{i} \pi \sum_{j=i+1}^{i+r-1} \hat{S}^z_{j} \right] \hat{S}^z_{i+r}
        \right\rangle \, ,
  \end{eqnarray}
  where the imaginary unit is $\mathrm{i}$.
  The string order parameter is an effective
  non-local operator to characterize hidden orders present in quantum
  phases of matter that cannot be described by the typical local
  operators~\cite{HaldanePLA1983, HaldanePRL1983, AltmanPRL2006, AltmanPRB2008}.
  The string order has shown signatures of
  of thermalization for scales related to the Lieb-Robinson bound~\cite{Mazza2014}. The
  remnant string order at finite times after a sudden quench out of the
  Haldane phase was credited to the preservation of symmetries of the
  Hamiltonian~\cite{Mazza2014}. However, if the symmetry is broken in
  the new phase after the sudden quench, then the string order is lost even
  at infinitesimal times in the thermodynamic limit. Such behavior
  makes the string order qualitatively different from the standard local
  order parameters~\cite{Strinati2016}. 
  In the case of finite
  systems, we measure $O_{i}(r=L-2i)$, where $i = 10$; this approach
  avoids boundary effects.
}
\item{The \emph{excess energy} measures the degree of excitation in
  the time-evolved states as:
  \begin{eqnarray}                                                                \label{eq:ResEnergy}
    \Delta E (\tau) = E_f -E_f^g \, ,
  \end{eqnarray}
  where $E_f=\langle \psi(\tau)| {\mathcal{H}}(\tau)|\psi(\tau) \rangle$, is the
  energy of the system described by $|\psi(\tau)\rangle$ after evolving through the quench 
  time $\tau$, and $E_f^g$ 
  is the ground state energy of the final Hamiltonian, $\mathcal{H}(\tau)$. The excess energy is equal to the weighted sum of all the
  excitation energies. As a result, this quantity will serve as the analog
  of the defect density originally considered by Kibble and
  Zurek~\cite{SantoroPRB2007, SantoroPRB2008, FazioPRB2008, DharPRA2015}.
  
}
\end{enumerate}

Some of the above quantities can be related to the speed of the quench in an universal manner through the well-known Kibble-Zurek mechanism.

This mechanism was originally proposed by Kibble in
the context of defects generated in the early universe \cite{Kibble1976},
which was later extended by Zurek to condensed matter
systems \cite{Zurek1985,ZurekPRL2005}.
This mechanism describes the formation of topological defects when the system is ramped
slowly across a second-order critical point, where the defect density depends
on the ramp rate exponentially. The exponents in such a dependence were shown
to be related to the universal equilibrium exponents of the underlying quantum phase transition.
Because of the divergence of the correlation length in the vicinity of the
critical point, it is impossible to ramp the system across the critical
point adiabatically without the formation of defects, thus signaling the
breakdown of the adiabatic theorem.

 Without loss of generality, we can consider a linear quench, such as 
 \begin{eqnarray}
  \epsilon(t) = \epsilon_0 + \frac{(\epsilon_f - \epsilon_0) \cdot t}{\tau} \, ,
  \qquad 0 \le t \le \tau \, ,
\end{eqnarray}
where $\epsilon$ is the quench parameter
as a function of time $t$, $\epsilon_0$ and $\epsilon_f$ are the initial
and final values of the parameter before and after the quench and $\tau$
is the time for quench. The parameter $\epsilon(t)$ can be $D$, $\theta$,
or a linear combination of both. Scaling analysis of the divergence of the correlation length $\xi$ shows that  any quantity which is related to the correlation length or the defect density, i.e., the excess
energy, 
will depend on the quench rate polynomially, with the exponent being a
combination of the critical exponents of the transition.  The defect density $n_{ex}$ follows the relation
\begin{eqnarray}                                                                \label{eq:KZpowerlaw}
  n_{ex} \sim \tau^{d\nu/(1+z\nu)} \, ,
  \label{eq:originalKZscaling}
\end{eqnarray}
where $d$ is the dimension of the system considered, $z$ and $\nu$ are
the critical exponents of the transition.  The scaling in Eq.~(\ref{eq:originalKZscaling}) has been
observed in many quench protocols, but it should also be noted that
there are systems where such a scaling analysis fails~\cite{DziarmagaAoP2010, MukundRMP2011}. For example,
there are some bosonic systems which remain non-adiabatic in the
thermodynamic limit. 

The scaling analysis of Eq.~(\ref{eq:originalKZscaling}) presumes the thermodynamic limit when the correlation 
length diverges to infinity at the critical point. However, in finite-sized systems, the 
maximum value of the correlation length can be the system size. Following a similar argument proposed
by the Kibble-Zurek mechanism, the minimum quench time needed for the system to attain 
adiabaticity is given by
\begin{eqnarray}
  \tau_{min}=L^{z\nu/(1+z\nu)} \, .
\end{eqnarray}
It is thus important to check for the
validity of the Kibble-Zurek mechanism in the BBM.  We do so by performing a linear quench across the phase transition and checking for estimates of the 
$\tau_{min}$ such that the final time-evolved state is close to the ground state in the Haldane phase.
Since the excess energy, $\Delta E$, is
a measure for the density of defects formed in the system, we can
use Eq.~\eqref{eq:originalKZscaling} by replacing $n_{ex}$ with
$\Delta E$ when checking for the validity of the Kibble-Zurek
mechanism. 

Quantities such as the Schmidt gap and entanglement entropy are related to the correlation length, and hence will scale with the quench time following the Kibble-Zurek mechanism as
\begin{eqnarray}
  \Delta \lambda \sim \tau^{-z\nu/(1+z\nu)} \, ,
  \label{eq:Schmidtgapscaling}
\end{eqnarray}

\begin{eqnarray}
  S = \frac{\mathcal{A}c\nu}{6(1+z\nu)} \log \tau + \mathrm{const} \, .
  \label{eq:entropyscaling}
\end{eqnarray}
For finite sized systems, a subsystem has two boundaries, and
hence ${\mathcal{A}=2}$~\cite{CalabreseJSM2004, ZurekPRA2007, VodolaPRB2014}.

\section{Quenching to the Haldane Phase                                                               \label{sec:results}}

\begin{figure}
    \centering
    \includegraphics[width=0.9\linewidth]{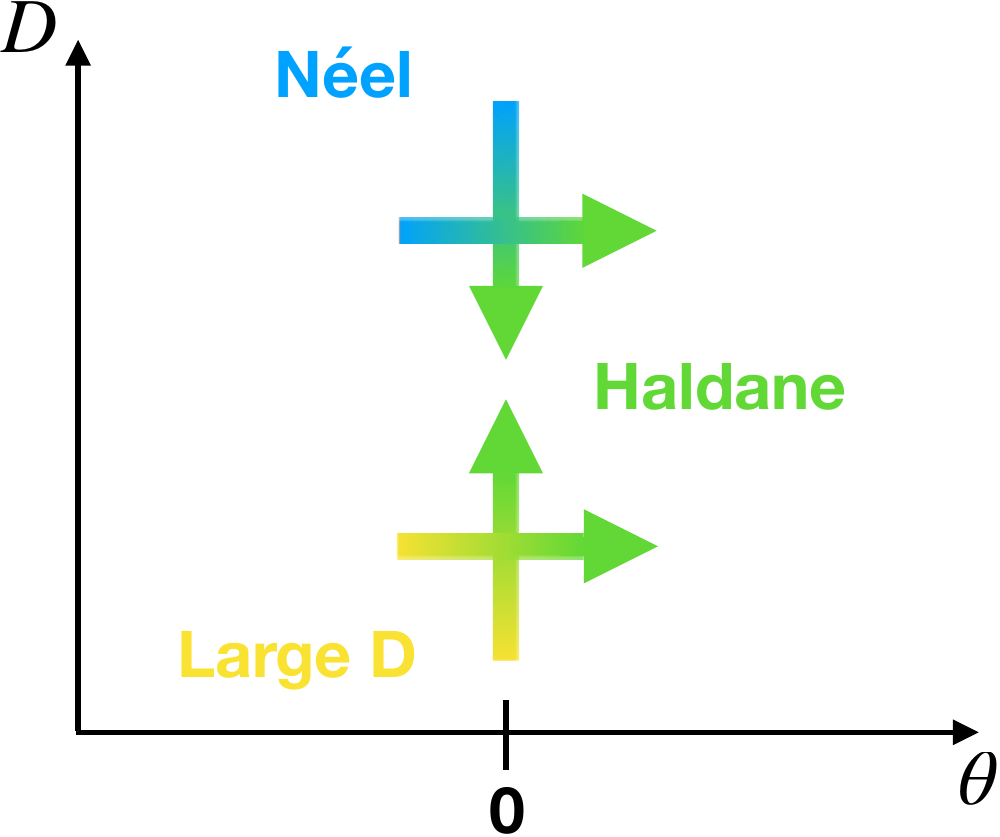}
    \caption{Schematic phase diagram: our studies encompass quenches of
      either the external field $D$ or the interaction
      $\theta$. The phase diagram contains the \Neel{} to
      Haldane phase transition, and the large-$D$ phase to Haldane
      transition. The exact values of the phase boundaries depend on finite-size effects
      and differ for the range of system sizes that we consider
      in our simulations -- a thorough study may be found in~\cite{DeChiara2011}.
    }
    \label{fig:sketch_theta0}
\end{figure}

In this work, we concentrate on reaching the Haldane phase from the \Neel{} phase and the large-$D$ phase
since these phases can be easily prepared in experiments with high fidelity. We
choose representative quantum critical points and perform a linear
quench across the respective quantum critical points as shown schematically in Fig.~\ref{fig:sketch_theta0}.

We begin the quench sufficiently far away from
the quantum critical point and terminate the quench
process approximately equidistant from the critical point on the other
side. The quench
time is then varied to study its effects on the quench processes. In addition, the system-size dependence
is studied by scaling up to systems as large as 200 sites. We categorize
the results according to the selected quantum phase transitions: subsections \ref{subsec:A}  and \ref{subsec:B} for linear quenches from \Neel{} and large $D$ to Haldane phase, respectively.

\begin{figure}[htbp]
  \begin{center}
  	  \includegraphics[width=0.95\linewidth]{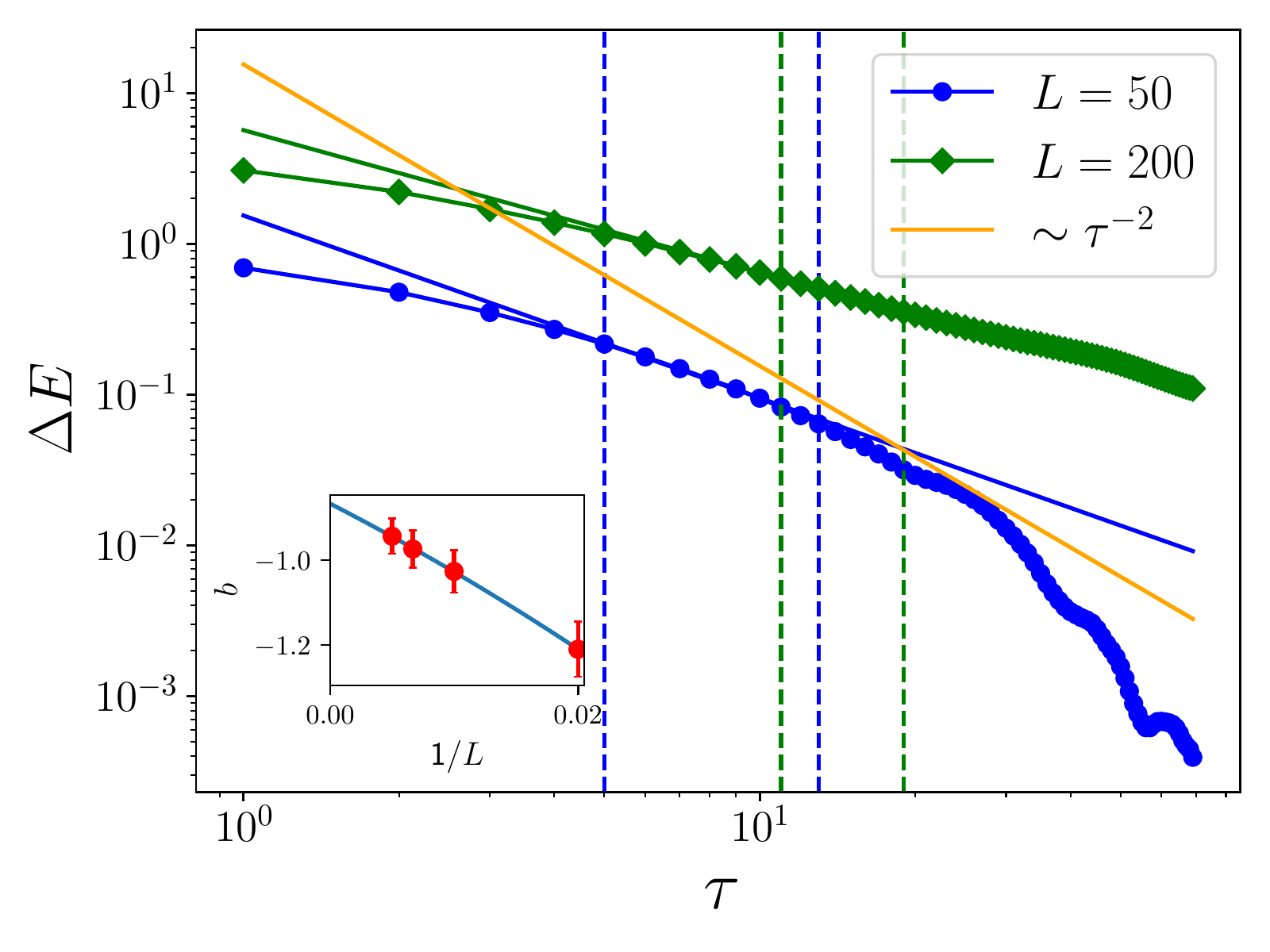}
    \caption{Excess energy $\Delta E$ as a function of quench times $\tau$ for two
      different lengths $L=50, 200$. We quench the uniaxial field from $D_i=-0.5$
      to $D_f=-0.15$; the interaction terms are constant at $\theta=0$. The dashed lines denote the
      region where the excess energy, $\Delta E$, is fitted by a power-law for the respective system sizes. The orange line denotes an
      inverse quadratic function, $\Delta E\sim \tau^{-2}$. (Inset) The values of Kibble-Zurek exponent $b(L)$, extracted
      by fitting a power-law to the excess energy in the intermediate $\tau$ values for different system sizes, $L$,
      plotted as a function of the inverse system size $1 / L$.
      A quadratic fit was employed to extrapolate
      an estimate for the thermodynamic limit, $b_\infty$.
    }
    \label{fig:ResEnergyPointA}
  \end{center}
\end{figure}

\subsection{\Neel{} to Haldane phase       \label{subsec:A}}

The \Neel{} phase is characterized by a finite non-zero spontaneous staggered
magnetization $M_z^{\mathrm{st}}$.
On the other hand, the Haldane phase represents a topological phase in
a 1-dimensional system signaled by a finite
string order defined in Eq.~\eqref{eq:StringOrder}. The Haldane phase
also displays a degeneracy in the entanglement spectrum, given by the
Schmidt gap as defined in Eq.~\eqref{eq:SchmidtGap}. Using a finite-size scaling of the staggered magnetization acting as the order parameter,
the phase boundary separating these two quantum phases has been
identified in an earlier work \cite{DeChiara2011}. Therein, finite-size scaling
analysis of the staggered magnetization and Schmidt gap yields the critical exponent $\nu=1.01$, consistent with the
Ising universality class, $\nu_{\mathrm{Ising}} =1$
\cite{DeChiara2011, SachdevQPT,CarrQPT}. For an Ising transition the critical exponent, $z=z_{\mathrm{Ising}}=1$ is known. However, for the present transition from \Neel{} to Haldane phase, the value of $z$ has not been calculated explicitly and numerically.

We perform a linear quench on the uniaxial field $D$ and keep the interaction $\theta$
fixed to $0$ during the time evolution.
We ensure that the initial and final values of $D$ are such that
they correspond to \Neel{} and Haldane phase, respectively, and are
far away from the quantum critical region.

We first look 
at the excess energy $\Delta E$ the system acquires
when quenched  across the quantum critical point, as defined in Eq.~\eqref{eq:ResEnergy}. $\Delta E$ is proportional
to the number of defects formed in the system after the quench,
and hence is expected to display Kibble-Zurek-like
scaling~\cite{RossiniJSM2009}. The behavior of $\Delta E$ as a function of the quench time $\tau$ is plotted
in Fig.~\ref{fig:ResEnergyPointA} for two  different system
sizes. Three distinct regimes can be seen from the plots. $\Delta E$ saturates to its maximum value for very fast
quenches, i.e., low values of $\tau$. This observation is consistent with
the fact that there can be a maximum number of excitations in the
system after the quench due to its finite size. The final state after the time evolution
is thus a superposition of several excited states. The value of
$\Delta E$ for the same value of $\tau$ increases
with the system size indicating a larger number of defects in the
system is formed after the quench for bigger systems. This
expected trend is due to the inverse dependence of the energy
gap on the system size at the critical point. For larger system
sizes, the gap is smaller; a smaller energy gap enhances the
probability of exciting the system to higher excited states,
which leads to a larger number of defects.

For very large values of $\tau$, applying the effective
Landau-Zener model would have resulted in a scaling of $\Delta E \sim \tau^{-2}$~\cite{ZurekPRL2005, RossiniJSM2009}. However, as shown
in Fig.~\ref{fig:ResEnergyPointA}, we do not quite observe
such a behavior. We could explore the Landau-Zener effect
for higher values of $\tau$ for the system sizes considered.
Such simulations are beyond the scope of the
current numerical techniques because of the large error involved.
In contrast, we observe the decay at large $\tau$ superimposed by
oscillations for $L=50$. These oscillations naturally arise when
effects of finite duration time are
considered~\cite{VitanovPRA1996, VitanovPRA1999, RossiniJSM2009}.
The frequency of the oscillations decrease with increasing system
size, along with a shift to higher quench times where the oscillations are
dominant. The oscillatory behavior can be suppressed by increasing the 
distance of the initial value $D_i$ from
the critical $D_c$ compared to the width of the critical regime~\cite{RossiniJSM2009}. The intriguing
quasi-adiabatic region lies between these two regimes where the
residual energy follows a power-law behavior. We attempt
to verify the Kibble-Zurek mechanism for this quench
protocol. For different lengths, we fit a power-law in the
intermediate power-law regime,
\begin{equation}                                                                \label{eq:powerlaw}
  \Delta E = a \tau^{-b(L)} \, .
\end{equation}

The value of the exponent, $b(L)$ calculated
through the exponential fitting procedure for different system sizes, $L$ is then fitted
using a quadratic scaling resulting in an
asymptotic value of $b_{\infty}\sim 0.866 \pm 0.008$ (see Fig.~\ref{fig:ResEnergyPointA}) for $L\rightarrow \infty$. The error bars include the errors arising from
the fitting procedure due to the range of $\tau$ values
used to fit and the fit itself. 

The phase transition from \Neel{}
to Haldane phase along the $D$ axis belongs to the Ising
universality class concluded from previous finite-size-scaling studies with the staggered magnetization and Schmidt gap~\cite{Sanpera_PRB2011, SanperaPRL2012}. 
We insert the corresponding critical exponents, $\nu_{\mathrm{Ising}}, z_{\mathrm{Ising}}$ in
Eq.~\eqref{eq:originalKZscaling} and obtain a value of $b_{\mathrm{theoretical}}=0.5$.
This number is quite different from the estimate extracted from our present
numerical calculations using excess energy $b_{\infty}$. The discrepancy can arise for two different reasons: either the Kibble-Zurek mechanism is not valid for this transition or the critical exponents are incorrect. Before coming to the conclusion that the Kibble-Zurek mechanism fails to describe the transition, let us try to verify all the critical exponents appearing in Eq.~\eqref{eq:originalKZscaling}. Since $\nu$ has been rigorously calculated using density matrix renormalization group (DMRG) method and Quantum Monte Carlo simulations, the only quantity that remains unknown is critical exponent $z$ since $d=1$. Assuming the Kibble-Zurek scaling to be valid, and inserting the value of $b_{\infty}$, we can estimate $z=0.159$. Using this value of $z$, we can see if other quantities such as the Schmidt gap and the entanglement entropy behave consistently according to the Kibble-Zurek mechanism with the obtained value of $z$. It should be noted that the possibility of anomalous Kibble-Zurek scaling exists as reported earlier for quenches across topological phases with edge states~\cite{Bermudez2009} which is robust to defect formation. However, that is unlikely the case here since we are quenching from a non-topological phase to a topological phase.

We point out that the reverse quench protocol, i.e.,
from the Haldane to the \Neel{} phase, finds
a similar power-law behavior of the excess energy in the intermediate regime of quench
times. We performed additional simulations to verify this observation, not shown here for brevity.

\begin{figure}
  \begin{center}
    \includegraphics[width=0.95\linewidth]{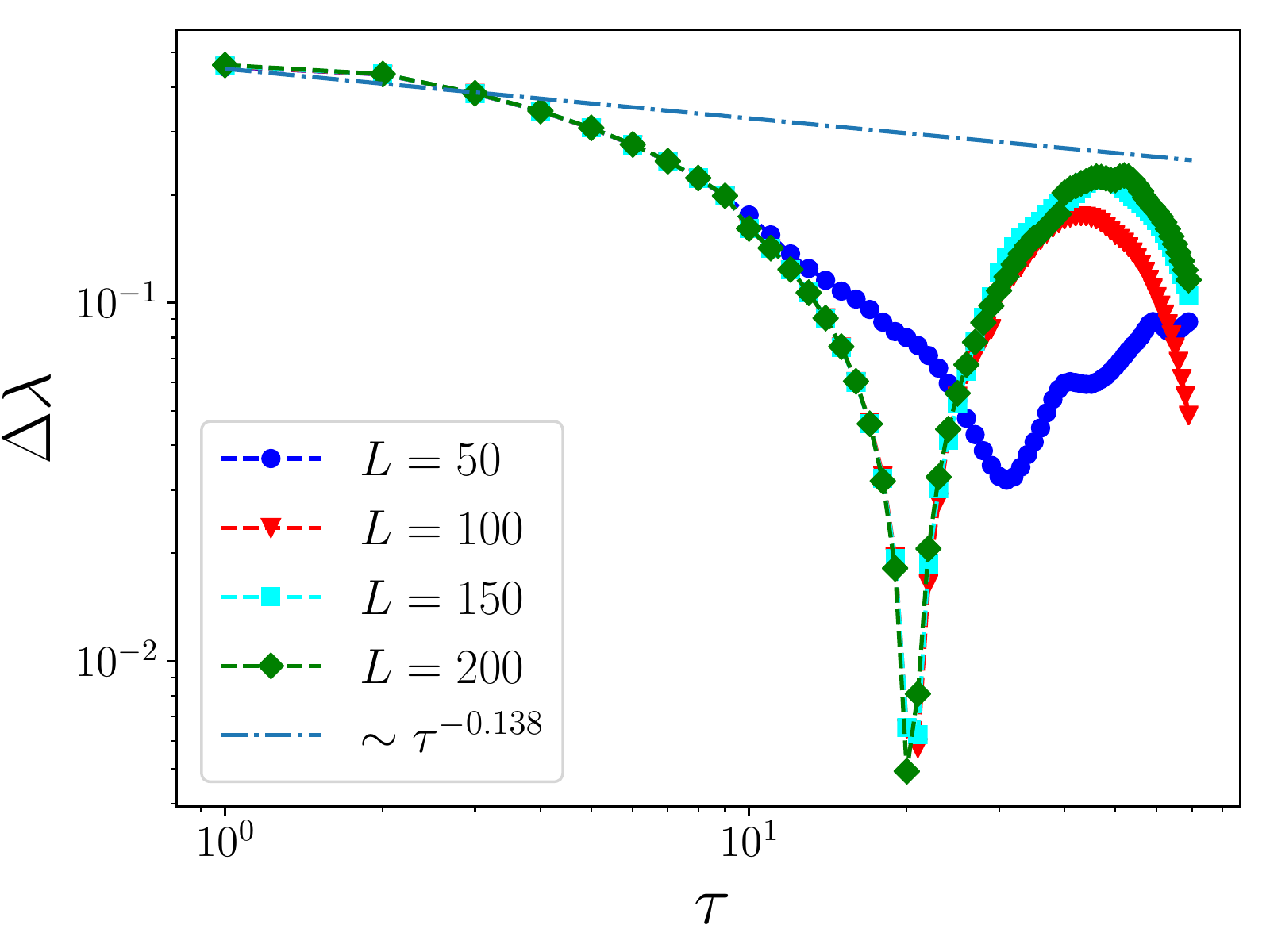} \\
    \includegraphics[width=0.95\linewidth]{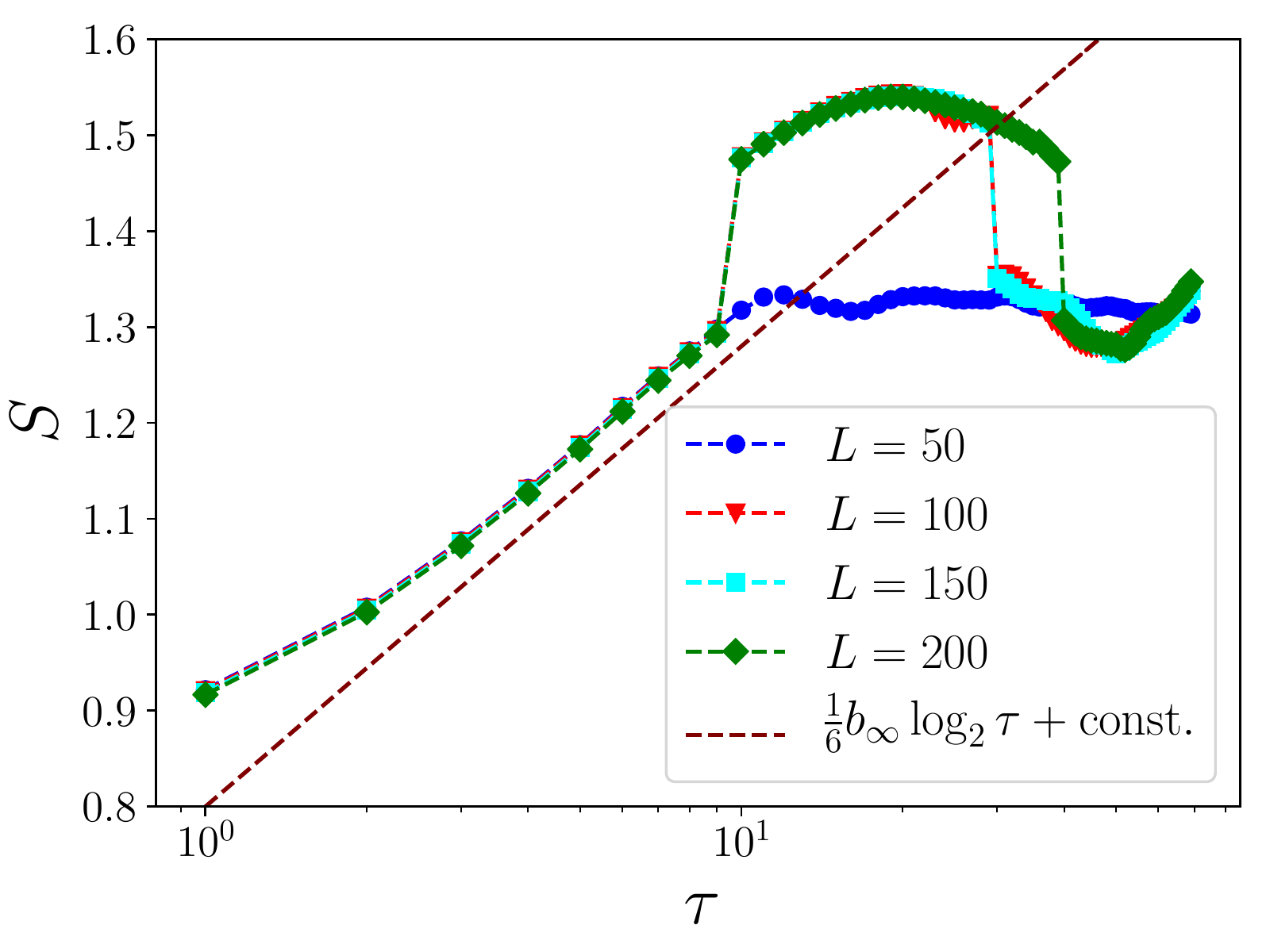}
    \caption{Schmidt gap $\Delta \lambda$ (top) as a function of the quench time $\tau$ for different lengths $L$  displays cusps related to the non-analytic behavior of the free-energy density. (Bottom) Block entropy, $S$ for as a function of $\tau$ for the same quench parameters as above. The maroon dashed
      curve in the plot of the block entropy corresponds to
      the Kibble-Zurek scaling from Eq.~\ref{eq:entropyscaling}. The quench protocol is similar to that shown in Fig.~\ref{fig:ResEnergyPointA}
    }
    \label{fig:SchmidtGapBlockEntropyPointA}
  \end{center}
\end{figure}

The Schmidt gap and the block entropy, as defined in
Eqs.~\eqref{eq:SchmidtGap} and~\eqref{eq:BlockEntropy}, are analyzed for the same quench protocol from \Neel{} to
Haldane phase by changing $D$, see Fig.~\ref{fig:SchmidtGapBlockEntropyPointA}.
Due to dependency of both quantities on the correlation length, they should also follow a Kibble-Zurek scaling.
Assuming the Kibble-Zurek scaling is valid for this transition, inserting the values obtained from the scaling of excess energy into Eq.~\eqref{eq:Schmidtgapscaling} leads to the following scaling of the Schmidt gap: 
\begin{eqnarray}
\Delta \lambda \sim \tau^{-0.138}  .
\label{eq:schmidtgapscalingNtoH}
\end{eqnarray}
Periodic oscillations are observed
for a given system size in the Schmidt gap related to crossings of
the first two eigenvalues in the entanglement spectrum, with an overall decay. This decay is power-law in nature with an exponent agreeing with Eq.~\eqref{eq:schmidtgapscalingNtoH}.
The cusps in the Schmidt gap are related to the periodic
non-analyticities of the free-energy density in the thermodynamic limit~\cite{KehreinPRL2013}.
The cusps are not an artifact of finite size effects since the position
of the cusps converge to one value for system sizes larger than $100$. A similar behavior of the Schmidt gap
have been reported earlier~\cite{HeylPRL2013, VodolaPRB2014} indicating the signature of a dynamical phase transition. 

The block entropy for small quench times, i.e., $\tau \leq 10$, coincides for all
the lengths considered in our simulations. This observation implies that the
entanglement spread has occurred on a length scale much smaller than any
of the system sizes considered in this work. Thus, we can set a lower limit
on the ramp velocities, i.e., the inverse of $\tau$, for which the size of the system
does not play an important role for the formation of entanglement.
For $\tau > 10$, the data for
$L=50$ deviates from the larger lengths:
the entropy abruptly jumps to a higher value. We note as expected that such
a separation for larger lengths is also observed in the Schmidt gap
behavior. The block entropies for the large lengths continue to
show a remarkable coincidence for significantly higher values of $\tau$.
The peak value at $\tau \approx 50$ in this regime occurs at the $\tau$ value for which we
observed the cusps in the Schmidt gap in Fig.~\ref{fig:SchmidtGapBlockEntropyPointA}. 

To check the validity of the
Kibble-Zurek mechanism, we insert $\nu=1.01$ and  $z=0.159$ in
Eq.~\eqref{eq:entropyscaling} with $c=1/2$~\cite{SanperaPRB2013,VodolaPRB2014}.
The logarithmic behavior is clearly seen for all system sizes for $\tau<10$ in Fig. \ref{fig:SchmidtGapBlockEntropyPointA}; thus, we confirm
the Kibble-Zurek behavior of the entanglement entropy. As expected, the
Kibble-Zurek prediction breaks down for smaller system sizes, i.e., $L=50$, and larger $\tau$ values.
There is an oscillatory behavior superimposed on the logarithmic
behavior which arises from the oscillatory nature of the entanglement
entropy. But to see the oscillations clearly, quenches with larger
$\tau$ values need to be simulated; these simulations are beyond the
scope of this work. Both the Schmidt gap and block entropy verify our initial assumption of the Kibble-Zurek mechanism to be valid, and hence the value of $z$ deduced from our calculations for this particular phase transition from \Neel{} to Haldane phase is proved to be correct.

\begin{figure}
  \begin{center}
    \includegraphics[width=0.95\linewidth]{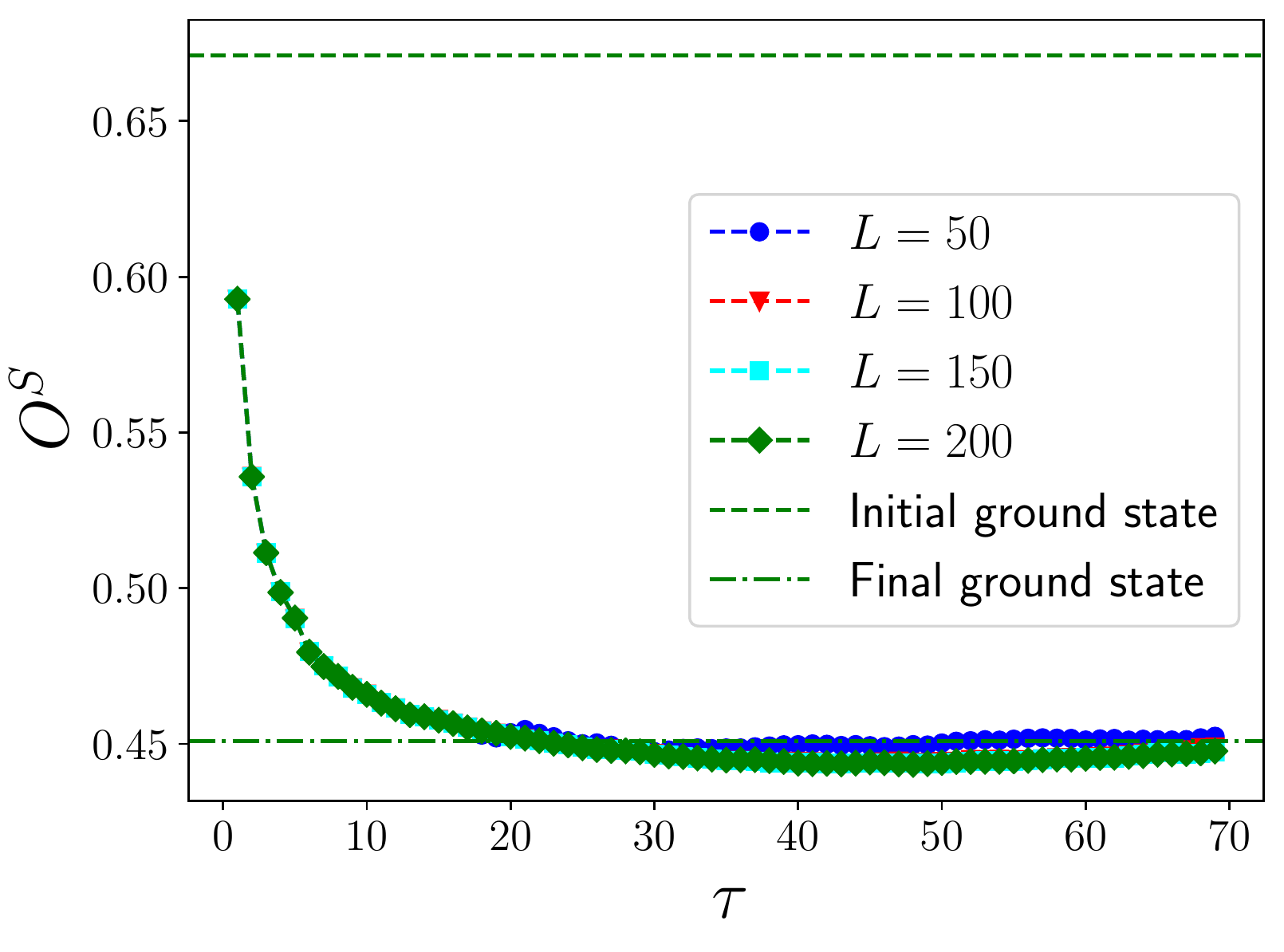}
    \caption{String order $O^S$ after a time evolution $\tau$ for
      different lengths $L$ decreases for slower quenches and becomes equal to the ground state value of the final Hamiltonian. The quench protocol is similar to that shown in Fig.~\ref{fig:ResEnergyPointA}. The initial (green dashed) and final (green dot-dashed) ground state values for the
      largest system size are shown for comparison.}
    \label{fig:StringPointA_Dchange}
\end{center}
\end{figure}

Both the initial and final phases display a finite string order, although
deep in the \Neel{} phase the string order is higher than in the Haldane
phase. Quenching from the \Neel{} to the Haldane phase, the time-evolved state
is expected to generate
defects since it crosses the quantum critical point. We consider it worthwhile
to study the fate of the string order after the quench in $D$ as shown in
Fig.~\ref{fig:StringPointA_Dchange}. We observe that for approximately
$\tau>20$,
the string order attains a value very close to that of the
ground state of the final Hamiltonian for all the system sizes. We can
thus conclude that the string order converges to the final ground state value very quickly
compared to the other observables shown before. This result indicates
the possibility of creating the Haldane insulator phase experimentally.
One needs to first prepare the \Neel{} phase, which has already been
achieved in experiments~\cite{MazurenkoNature2017}, and then quench $D$ with $\tau>20$
to reach the Haldane insulator phase for system sizes considered here.
The quench process will not excite the system enough to kill the string
order, as reported in earlier theoretical works with sudden
quenches~\cite{FazioPRB2014, FazioPRB2016}.

\begin{figure}
  \begin{center}
    \includegraphics[width=0.95\linewidth]{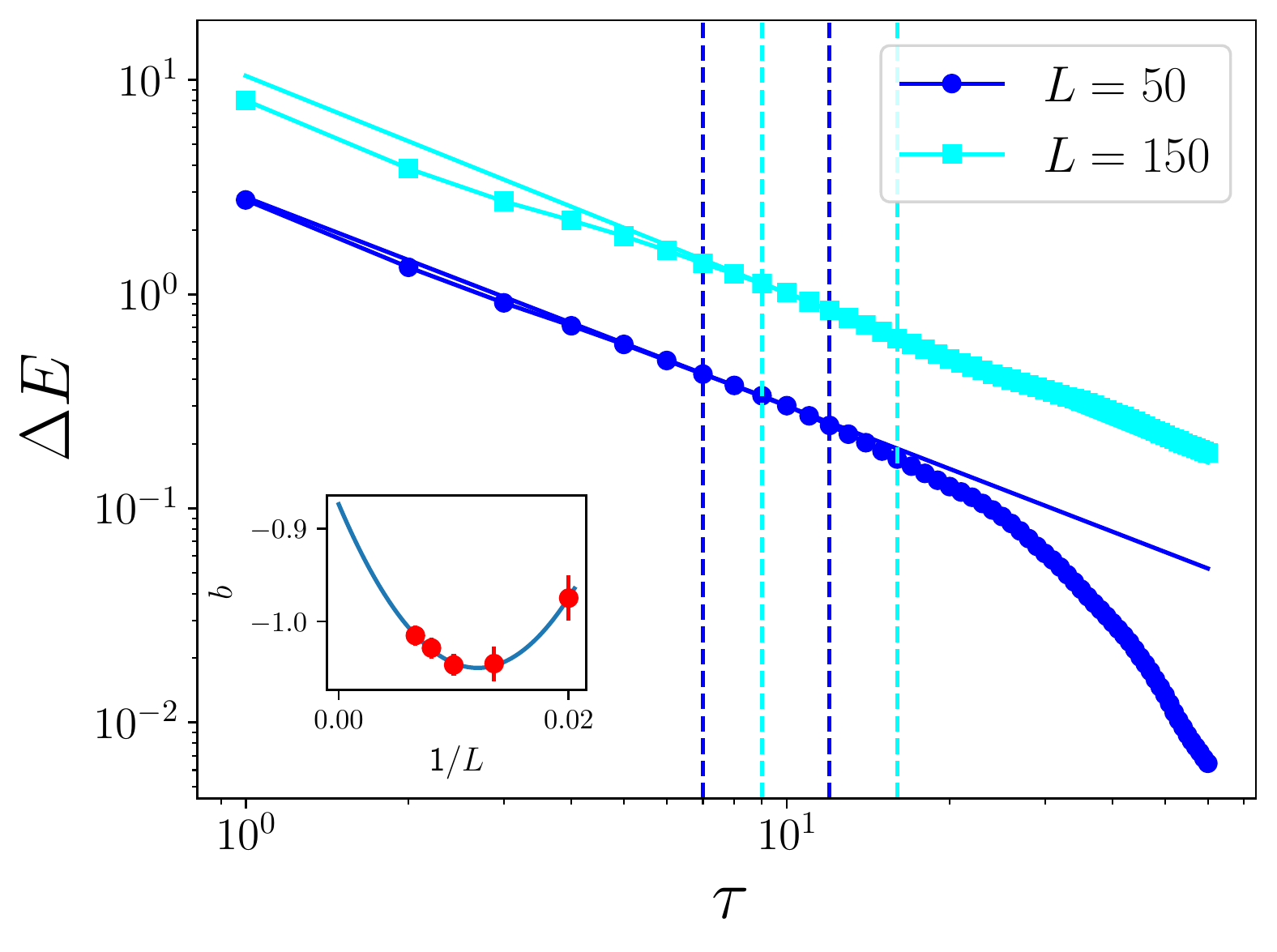} \\
    \includegraphics[width=0.95\linewidth]{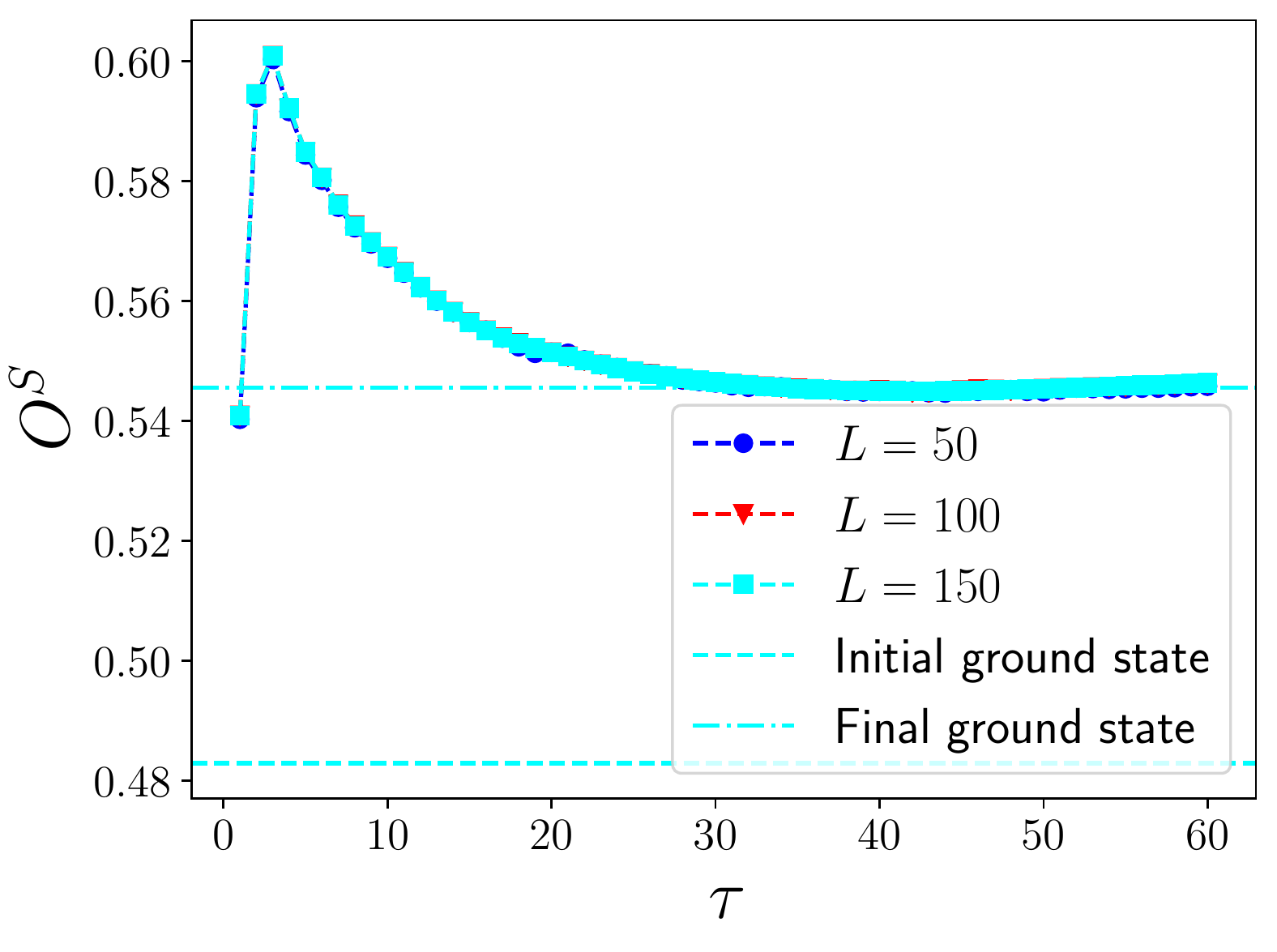}
    \caption{Top: Quenching from Haldane to \Neel{} phase by changing
      the tuning of the interaction $\theta$
      from $-0.1\pi$ to $0.1\pi$ at a constant uniaxial field of $D=-0.310$
      exhibits a different behavior
      for excess energy $\Delta E$ with a power-law behavior denoted by the red line with the power-law exponent, $b=0.873\pm 0.011$ for a system size of $L=150$. Bottom: String order $O^{S}$ after the linear quench  for intermediate quench times becomes higher than the final ground state value. The initial (dashed cyan line) and final (dot-dashed cyan line)
      ground state values of the string order are shown for comparison.                   \label{fig:ResEnergyStringPointAThetachange}}
  \end{center}
\end{figure}

The quench from the \Neel{} to the Haldane phase can also be achieved along the
$\theta$ direction. 
Figure~\ref{fig:ResEnergyStringPointAThetachange}
shows the behavior of the excess energy as a function of the quench
time after a quench by changing $\theta$, holding $D$ constant at $-0.310$. We find a power-law region for intermediate $\tau$ values. Extrapolating the power-law exponents for different system sizes gives a value of $b_\infty=0.873\pm 0.011$, which is consistent with the extrapolated value obtained in the previous protocol of quenching the external field from \Neel{} to Haldane phase, where we found $b_\infty=0.866\pm 0.008$.

We now look at the evolution of the string order parameter, $O^S$ after
the quench in the $\theta$-direction in Fig.~\ref{fig:ResEnergyStringPointAThetachange}. Before proceeding to study the effect of the quench, it is informative to notice the ground state values. Typically the \Neel{} phase has a larger string order compared to the Haldane phase. This observation is always true when the interaction, $\theta$ is fixed and external field, $D$ is varied. However, this statement may not be true when going from \Neel{} to Haldane phase through other trajectories in the phase diagram. For the protocol where $D$ is kept fixed and $\theta$ changed, we clearly see that the ground state string order in the \Neel{} phase is smaller than in the Haldane phase.  Following the linear quench,  the string order converges to the ground state value of the final Hamiltonian for longer quench times, see $\tau > 30$. In contrast, we observe that for very short quench times the string order increase monotonically, till it reaches a maximum at $\tau \sim 3$. Surprisingly, for intermediate quench times ($2<\tau<30$), the string order  after the time evolution is larger than the final ground state value. This behavior is indeed not expected, and it implies that for a certain range of quench times, the excitations are produced in such a way that enhances the final string order. This observation implies clearly that using this quench protocol the ground state of the Haldane phase may not be reached for intermediate times, but the final state will have larger string order, offering a surprising and useful experimental prescription to maximize string order.

\subsection{Non-Haldane (large-D) to Haldane phase   \label{subsec:B}}

We now investigate the effects of quenching from the large $D$ phase to the Haldane phase 
as shown in the schematic Fig.~\ref{fig:sketch_theta0}. We follow a similar approach as before, quenching by changing $D$ and $\theta$. Previous studies~\cite{SanperaPRB2013} have shown that this transition has a critical exponent $\nu=1.56$ and a central charge $c=1$, both distinct from the values of $\nu$ and $c$ corresponding to the transition from \Neel{} to Haldane phase.

\begin{figure}
  \begin{center}
    \begin{tabular}{cc}
      \includegraphics[width=0.9\linewidth]{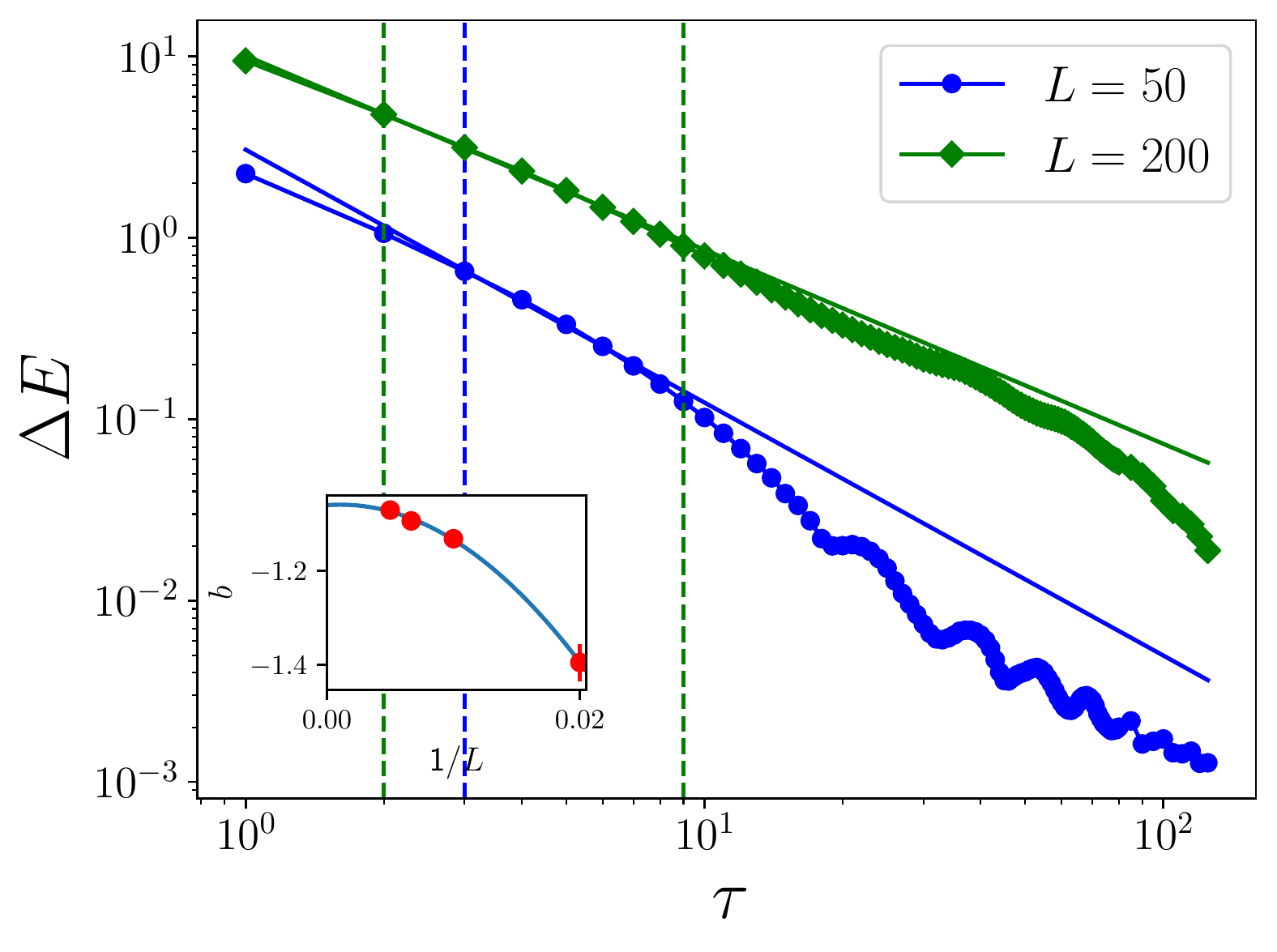}
    \end{tabular}
    \caption{Excess energy $\Delta E$ displays a power law behavior as a function of quench time $\tau$, shown here for two
      different lengths $L$ after quenching from a uniaxial field $D_i=1.6$ to $D_f=0.6$;
      the interaction is held constant, $\theta=0$. The vertical dashed lines denote the region
      which is fitted by the power-law: $\Delta E= a\tau^b(L)$, for the corresponding system size, $L$. (Inset) Quadratic extrapolation of $b$-values for different system sizes, $L$, yields $b_\infty = -1.06 \pm 0.021$.}
    \label{fig:ResEnergyPointC_DChange}
  \end{center}
\end{figure}

Figure~\ref{fig:ResEnergyPointC_DChange} clearly shows a power-law
region as expected from the Kibble-Zurek mechanism.  Characteristic oscillations in the residual energy are observed for very slow quenches. Due to computational complexity, this feature is visible only for smaller lengths. Fitting the extracted value of the power-law exponent $b(L)$ in the quasi-adiabatic regime with a quadratic function of the system size, we obtain $b_\infty = -1.06 \pm 0.02$. This value is certainly different than what was obtained in the previous quench protocol of \Neel{} to Haldane phase. Since the Kibble-Zurek mechanism is valid, we can extract the critical exponent $z$ as before. This analysis yields $z=0.229\pm 0.004$.

\begin{figure}
  \begin{center}
    \includegraphics[width=0.9\linewidth]{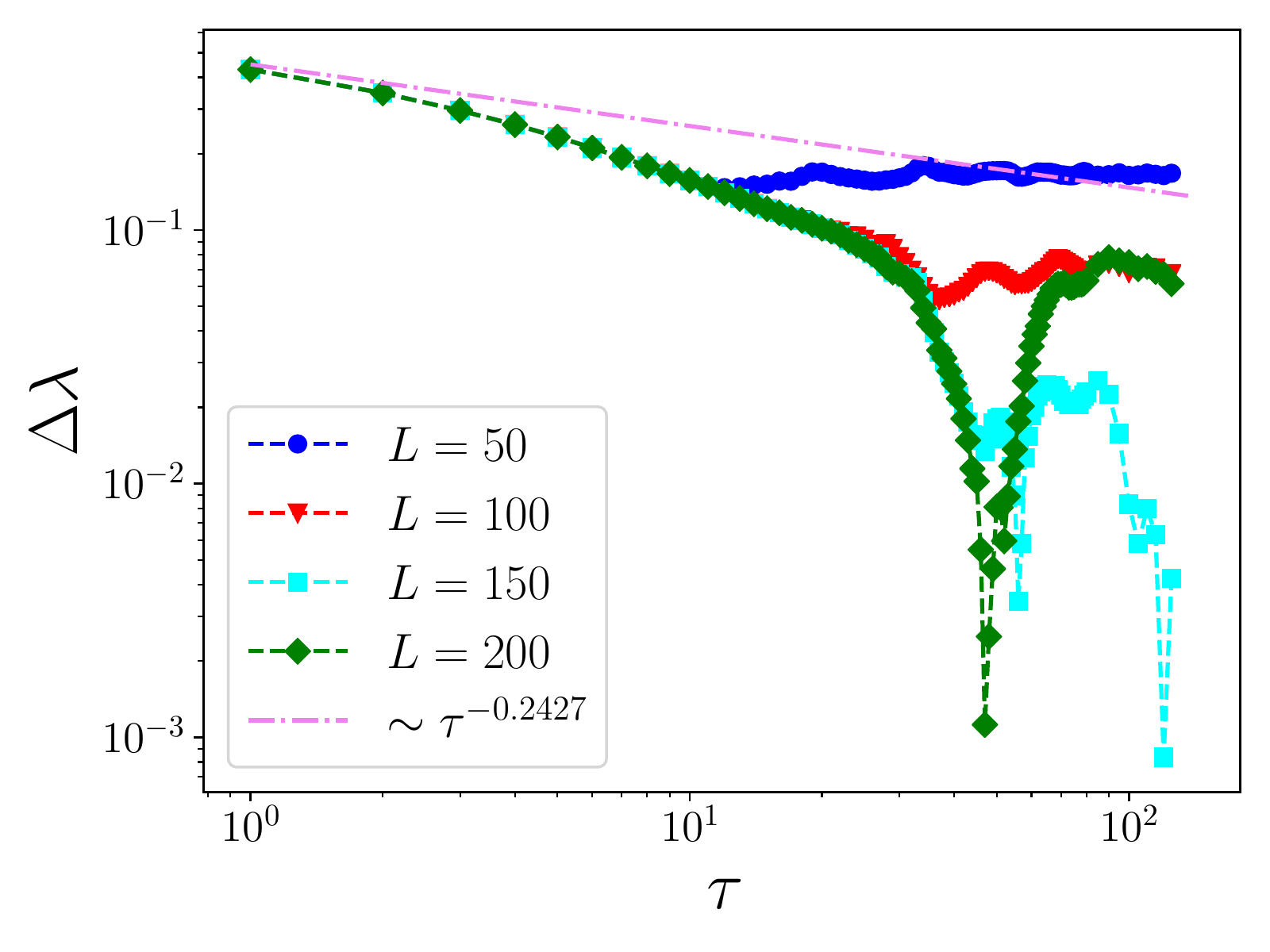} \\
    \includegraphics[width=0.9\linewidth]{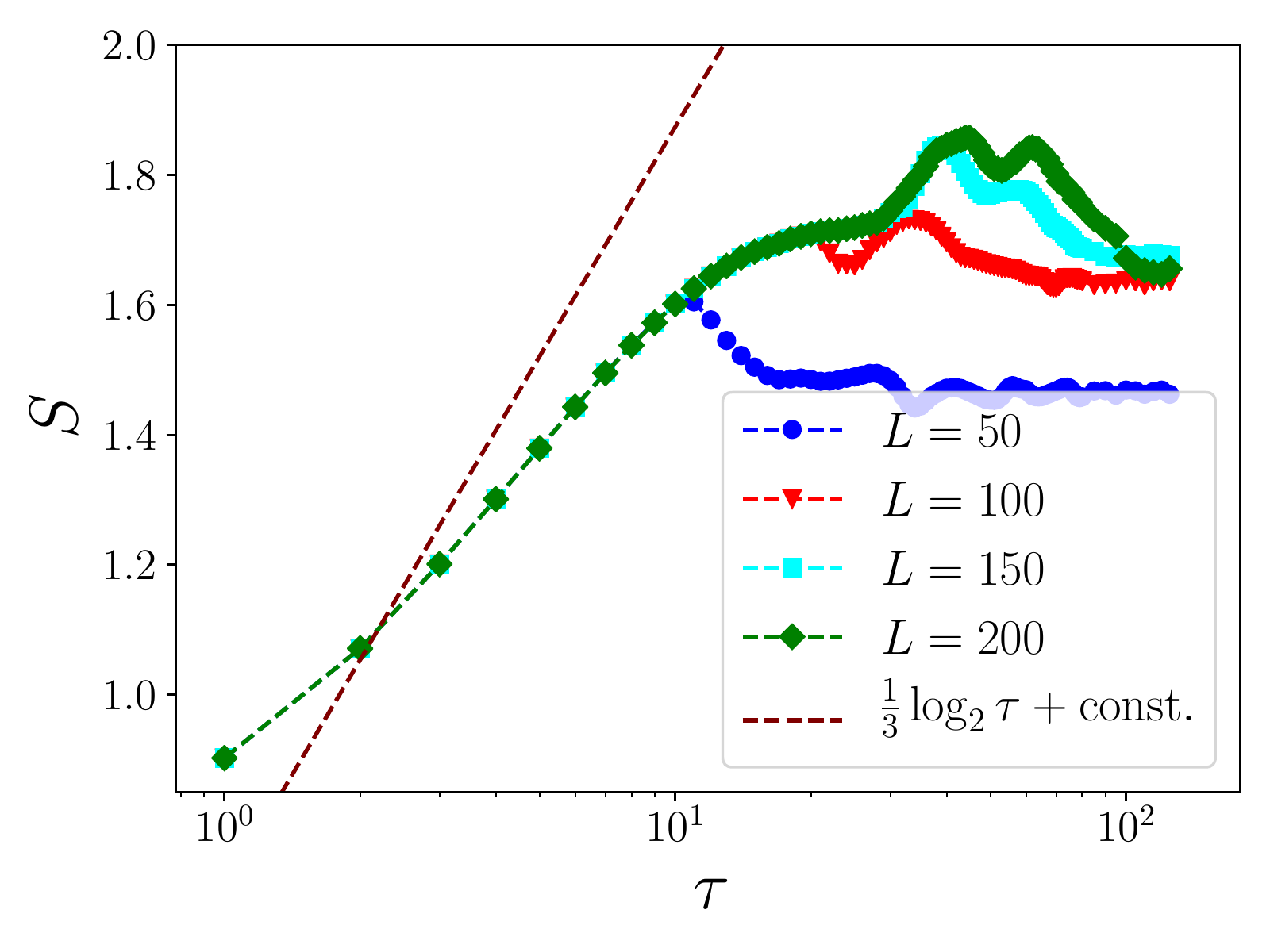}
    \caption{(Top) Schmidt gap after the quench has cusps at different quench times for larger system sizes. The pink dot-dashed line is the expected power law envelop from Eq.~\ref{eq:Schmidtgapscaling}. (Bottom) The block entropy shows a marked deviation from the expected Kibble-Zurek scaling, shown by the maroon dashed line. The quench protocol is similar to that mentioned in Fig.~\ref{fig:ResEnergyPointC_DChange}.}
    \label{fig:SchmidtGapBlockEntropyPointC_Dchange}
  \end{center}
\end{figure}

The final Schmidt gap displays cusp-like behaviors arising from
the non-analyticity, but only for larger lengths, i.e., $L=150,200$,
which can be early-time indications of dynamical phase transitions
as shown in Fig.~\ref{fig:SchmidtGapBlockEntropyPointC_Dchange}. If
the quench is done rapidly ($\tau<8$), the largest two eigenvalues
of the reduced density matrix remain almost degenerate. However for $L>150$, the Schmidt gap is two orders of magnitude lower than
that in the adiabatic limit ($\tau > 20$) when compared to smaller systems of $L<100$.
Due to the fact that the Schmidt gap in the final ground state is smaller than
its initial ground state value, the oscillations exhibit a power-law decay envelope. 

The entanglement entropy after the time evolution shows a
characteristic power-law behavior for smaller quench times,
followed by oscillations. Fitting this power-law region with Eq.~\eqref{eq:entropyscaling}, the corresponding values for this transition do not show a good agreement. This behavior can imply either the Kibble-Zurek mechanism is not valid in this transition, and hence the $z$ value derived is incorrect, or the entanglement entropy does not comply with the Kibble-Zurek mechanism. Deviations from the power-law behavior occur progressively at larger
quench times with increasing system size.  The
oscillations typically occur when the power-law region ends,
and can be attributed to the presence of excited components in 
the wave function after crossing the critical
point~\cite{VodolaPRB2014}.

\begin{figure}
  \begin{center}
    \includegraphics[width=0.9\linewidth]{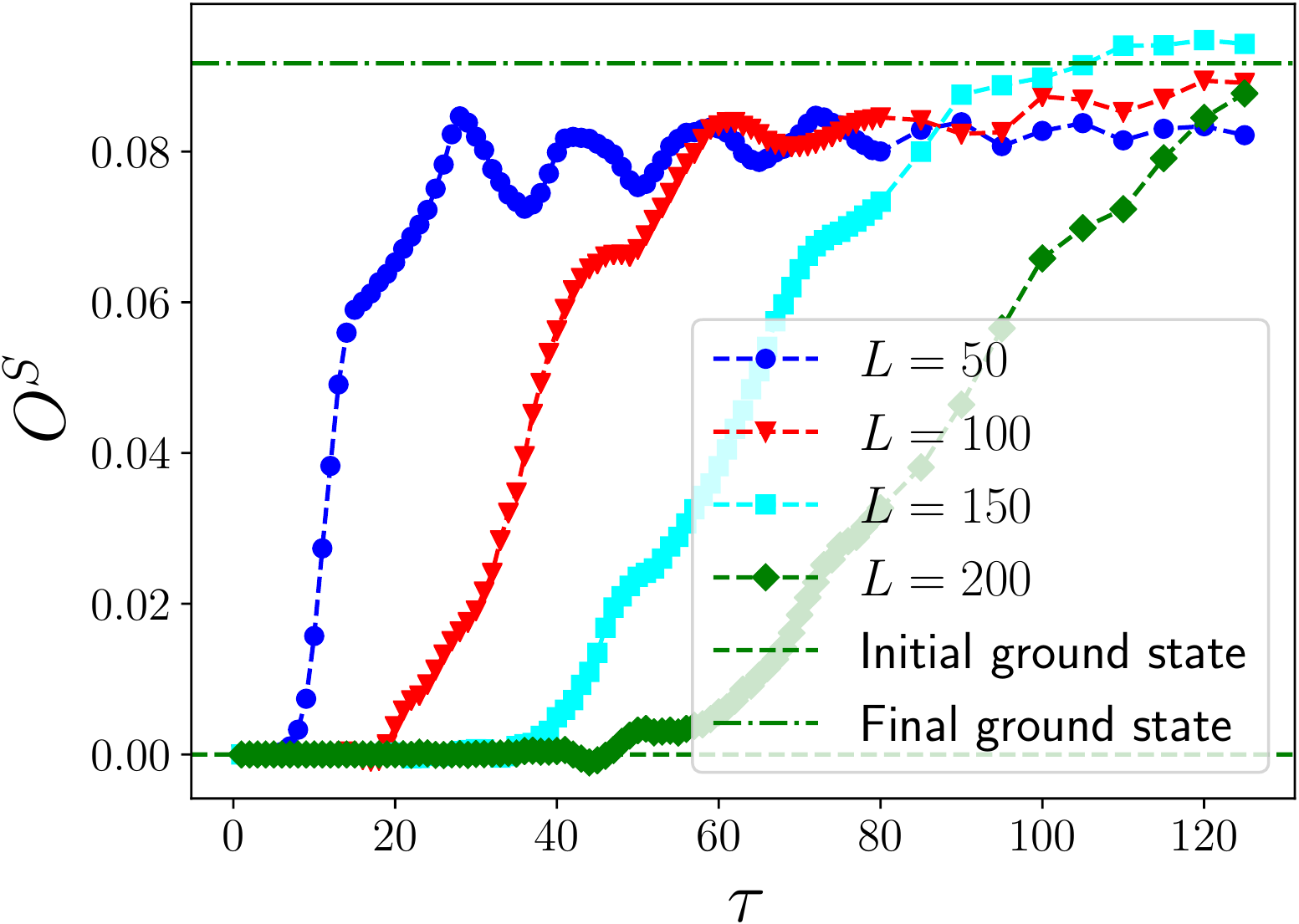}
    \caption{Following a quench as mentioned in Fig.~\ref{fig:ResEnergyPointC_DChange}, the string order increases towards the final ground state value for slower quenches. However, larger systems require longer quench times to reach the final ground state value, shown by the green dot-dashed line.}
    \label{fig:StringPointC_Dchange}
  \end{center}
\end{figure}

The Haldane phase has a finite string order unlike the
large $D$ phase. Following a quench from the large $D$ to
Haldane phase, the string order is found to approach the final ground state value in the Haldane phase for all system sizes as
shown in Fig.~\ref{fig:StringPointC_Dchange}. Oscillations in the time-evolved string order as a function of the quench time in the adiabatic limit is noticeable for smaller system sizes whereas the proximity of the converged time-evolved string order to the final ground state value increases as the system size increases. Both these effects can be attributed to finite-size effects. The quench time at which the evolved string order converges close to the final ground state value depends explicitly on the system size. 

\begin{figure}
  \begin{center}
    \begin{tabular}{cc}
      \includegraphics[width=0.9\linewidth]{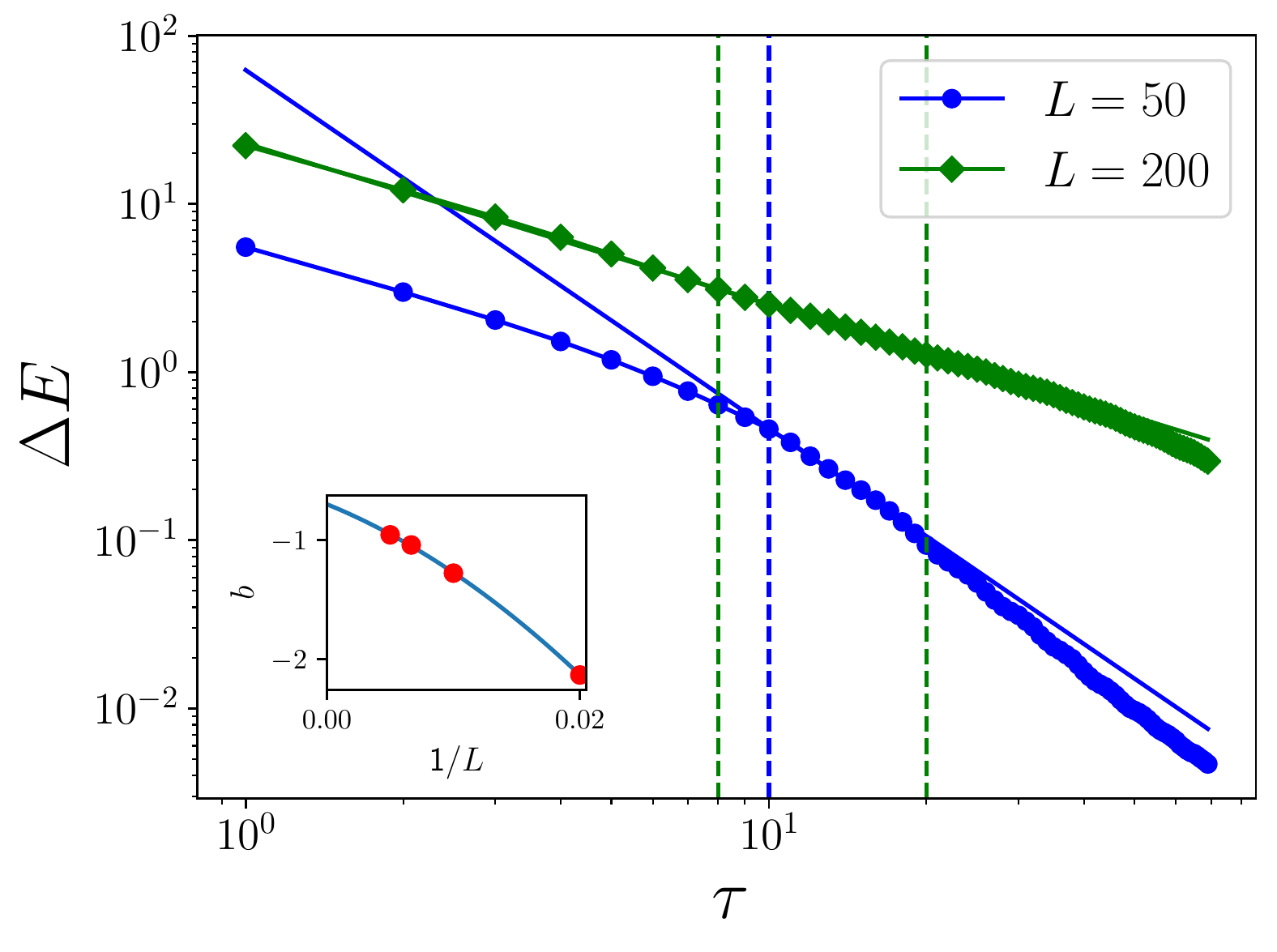} 
    \end{tabular}
    \caption{Excess energy $\Delta E$ as a function of quench times $\tau$ for two different
      lengths $L$ after quenching the interacting parameter from $\theta_i=-0.25\pi$ (Large $D$ phase)
      to $\theta_f=0.15\pi$ (Haldane phase); the uniaxial field is held constant at
      $D=0.5$. The dashed line denotes the region which is fitted
      by the power-law: $\Delta E= a\tau^b$. (Inset) The thermodynamic value of $b_\infty=0.701\pm 0.001$ is obtained when $b(L)$ is fitted with a quadratic function of $1/L$. }
    \label{fig:ResEnergyPointC_Tchange}
  \end{center}
\end{figure}

\begin{figure}
  \begin{center}
    \includegraphics[width=0.9\linewidth]{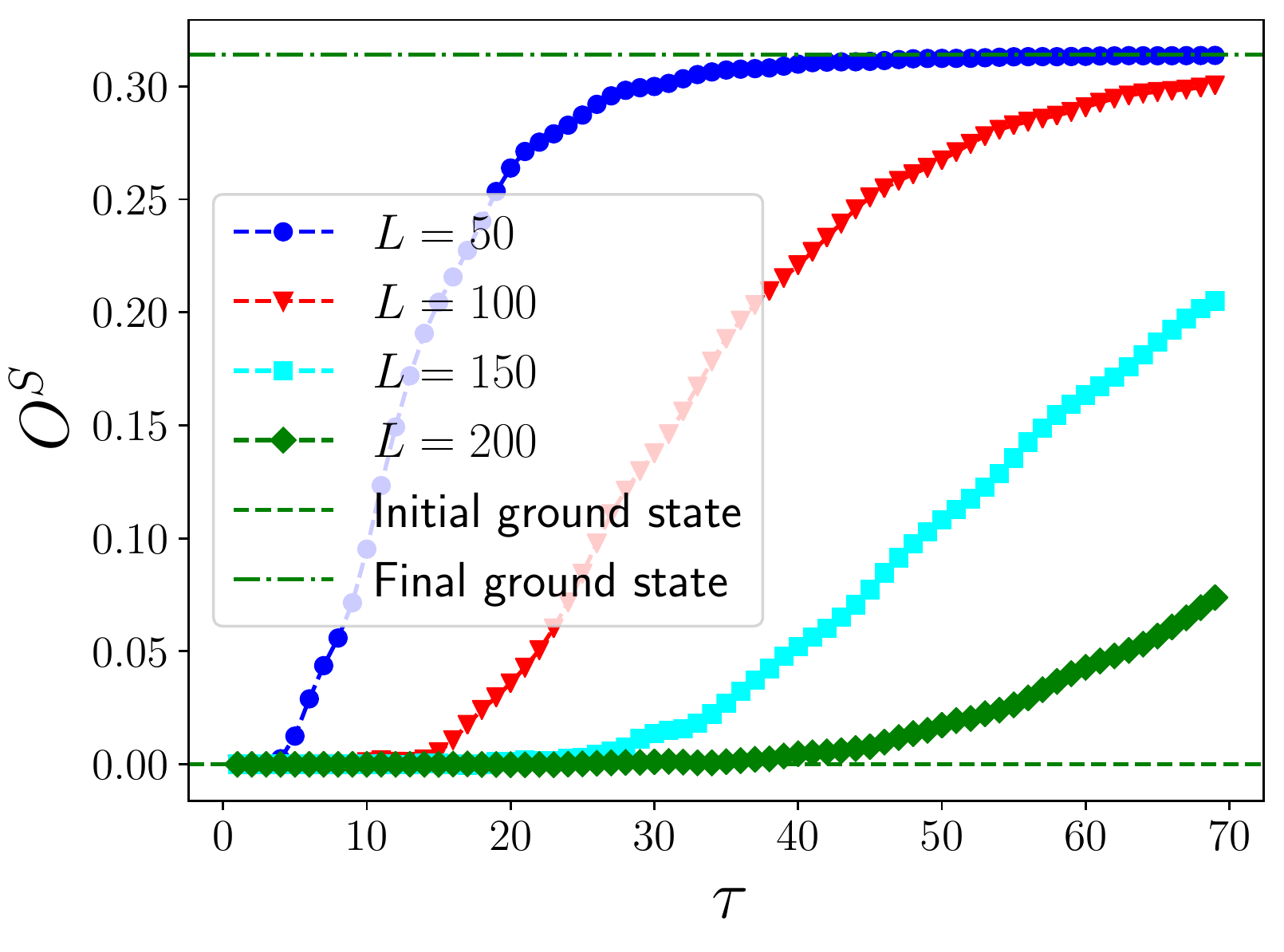}
    \caption{String order $O^{S}$ after the time evolution for different lengths $L$
      corresponding to the quench from $\theta_i=-0.25\pi$ to $\theta_f=0.15\pi$ while
      holding $D=0.5$. }
    \label{fig:StringPointC_Tchange}
  \end{center}
\end{figure}

The quench from large $D$ to Haldane phase can also be attained by
changing $\theta$ and keeping $D$ fixed.
Figure~\ref{fig:ResEnergyPointC_Tchange} shows the behavior of the
excess energy as a function of quench times. Extracting the power-law exponent from the quasi-adiabatic regime and extrapolating to the thermodynamic limit suggests a value of $b_\infty=0.701\pm 0.001$. This value is markedly different from that obtained when quenching along the external field, $D$, from the large $D$ phase to Haldane phase, suggesting a different universality class of the transition depending on the path taken from one phase to the other. Examining the string order after the quench in Fig.~\ref{fig:StringPointC_Tchange}, we recognize a system-size-dependent behavior. The evolved string order reaches the final ground state value at quench times which increases with system size. Unlike the previous quench protocol, we observe neither any oscillations in the time-evolved string order nor a size-dependent difference  between the ground state and the evolved string order. However, it should be noted that both the quench protocols show a system-size-dependent behavior of the string order when evolving from large $D$ to Haldane phase. This observation is in stark contrast to the scenario when evolving from the \Neel{} to the Haldane phase, where the string order has no dependence on the system size. Such a behavior will be very useful for experimental groups where observations are likely to show finite-size effects.

\section{Methods and error analysis                                                               \label{sec:methods}}

We used two different MPS packages to simulate statics and dynamics of the
Hamiltonian described in Eq.~\eqref{eq:HBBM}, i.e.,
openTEBD \cite{OpenTEBDPackage} and
openMPS \cite{Jaschke2018,OpenMPS}. The ground state simulations were carried out with bond dimension 1000, and a maximum system size of 200 lattice sites. For the time-dependent simulations, fourth order Suzuki-Trotter decomposition was used with typical time steps of $0.01$, truncation error limit of $10^{-10}$ and bond dimensions between $500$ to $800$ depending on the quench protocol and system size.


We  support our results from the previous section with the following error
analysis. Numerical simulations with tensor network methods have two
main sources of error. On the one hand, we have the Trotter approximation
scaling as a power of the time step $dt$ \cite{Schollwoeck2011}. On the other hand, we have the
truncation of the Hilbert space in terms of the bond dimension. Both errors
are applicable to the statics and dynamics as we use imaginary time
evolution for the statics; a variational ground state does not have a
Trotter error~\cite{Carr_NJP2009}. Error bounds can be calculated
for observables based on the variance of the energy and the energy gap
to the first excited state for the ground state \cite{Jaschke2018}.
The $n^{\mathrm{th}}$ order Trotter decomposition has
a well-controlled error proportional to $\mathcal{O}(dt^{n})$. Thus, we
concentrate in the following on the bond dimension, where the maximum grows
exponentially with the system size. The validity of a truncation to a
value well below the maximum is demonstrated in the following.

\begin{figure}
    \centering
    \includegraphics[width=0.9\linewidth]{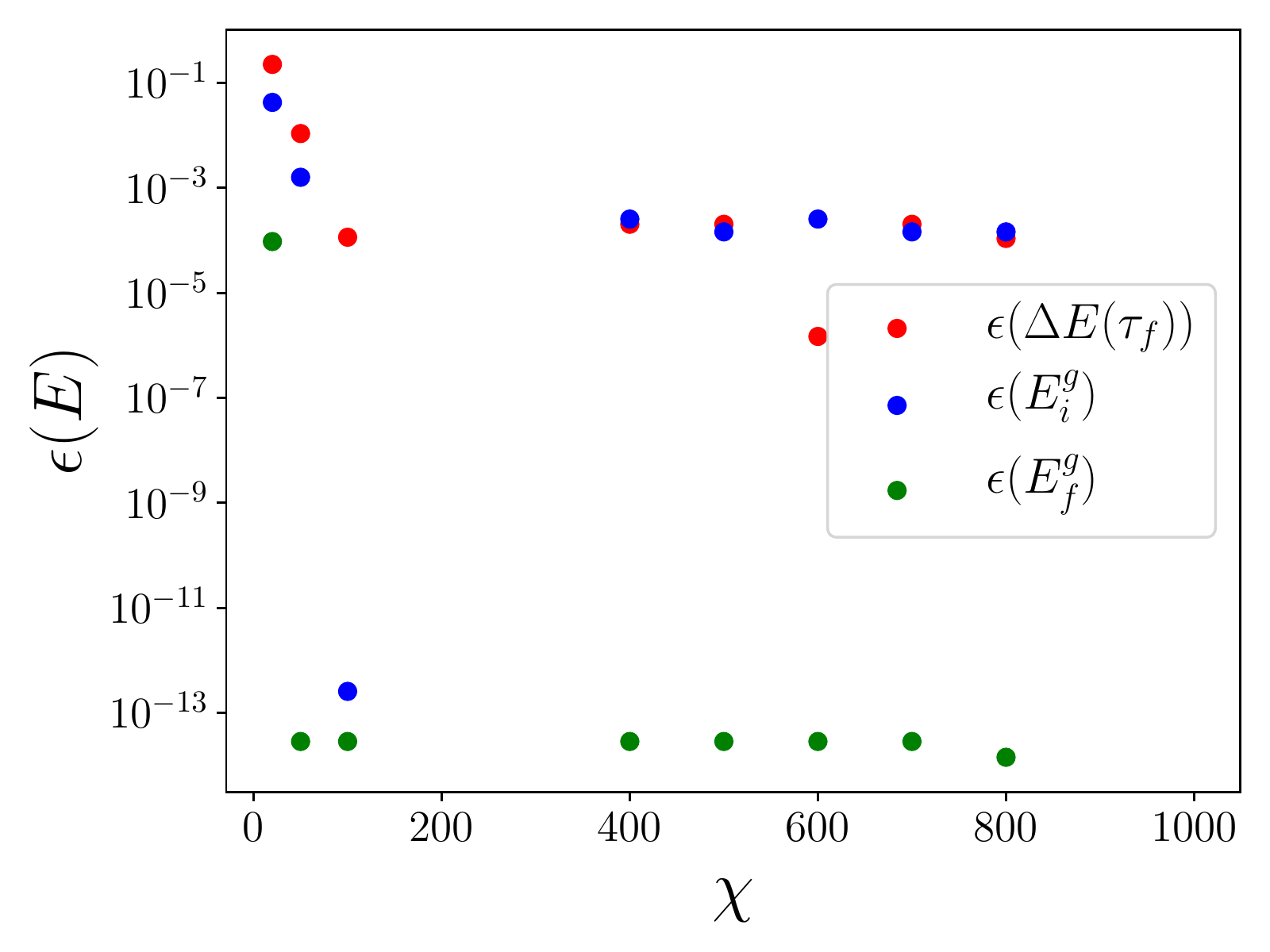}
    \caption{Error analysis: The difference in the final energy after time evolution, 
      the initial ground state energies and the final ground state energies for 
      different bond dimensions, $\chi$, with the corresponding value obtained using the
      largest bond dimension,  as a function of bond dimensions for 
      Haldane phase to large $D$ phase quenches and a system size of 200 sites.
    }
    \label{fig:errorfig}
\end{figure}

Our result for the Kibble-Zurek scaling are based primarily on the excess energy.
Three states emerge in the error analysis: (i)~the ground state serving
as the initial state to the quench; (ii)~the ground state of the final
parameters of the quench; and (iii)~the final state of the quench.
These three values are sufficient to track down the main source of error.
Figure~\ref{fig:errorfig} shows the convergence of the energy measurements,
which are the foundation of the excess energy. We varied the bond
dimension from $25$ to $1000$. In addition, we show the difference
from the simulation with the highest bond dimension. The convergence
for the simulated bond dimension is straightforward. The remaining error of the order of $10^{-15}$ is an artifact of the machine precision in numerical simulations. This short error analysis justifies us in dismissing error bars from the numerical simulations in the fitting procedure for the Kibble-Zurek
scaling.

\section{Conclusion                                                             \label{sec:concl}}

We have examined the effects of linear quenches across quantum 
critical points in the BBM with a quadratic Zeeman 
field. Our primary focus has been to quench into the Haldane phase from phases which can be readily prepared in experiments, such as the \Neel{} and large $D$ phases. We evaluated the validity the of Kibble-Zurek mechanism in each of these transitions by 
studying excess energy and different observables which are related to the correlation length as a function of quench time, such as the block entropy and the Schmidt gap.
We base our results on numerical simulations with matrix product states methods
in 1-dimensional finite-size systems.

For the \Neel{} to Haldane phase transition, we found Kibble-Zurek like 
power-law behavior in the excess energy. However, the behavior of the excess
energy, along with the power-law exponent, depends on the quench protocol, 
i.e., whether we perform the quench along external magnetic field, $D$, or interaction parameter, $\theta$. The Schmidt gap
and block entropy shows signs of dynamical phase transitions through the
appearance of cusps at specific quench times. The power-law exponent derived from the behavior of excess energy follows the Kibble-Zurek mechanism if the correct critical exponents are considered. 

We find that the behavior of string order 
depends heavily on the quench protocol, with a marked difference in the
two scenarios. Quenching across $\theta$ reveals a regime for quench times in which the final time-evolved string order is larger than the final ground-state string order, suggesting the formation of defects due to the quench which not only preserves the string order, but also enhances it. Furthermore, the behavior of string order as a function of quench times shows almost no dependence on the system size.  Quenching from the large $D$ to the Haldane phase shows yet again the dependence on the quench direction in the phase diagram. The string order shows a clear system-size dependence, with larger systems needing longer times for the final time-evolved string order to become approximately equal to the final ground-state value. Our observations of the string order following a linear quench across the two quantum phase transitions will play an important role in experiments in search of the elusive Haldane phase.

We have seen that the investigation of the transitions to the
Haldane phase yields intriguing physics. These results raise the question
of which other phenomena the remaining phase transitions in the
BBM may contain. These transitions are 
a fruitful subject for future studies. Secondly, it would be important to
eventually model the system as an open systems coupled to a reservoir, e.g., a with
a Lindblad master equation in order to examine the question of decoherence in quantum simulators seeking to identify and explore the Haldane phase. Finally, a remaining open question is which cause for defects, i.e.,
the quench rate or the open system effects, is predominant for a given
parameterization of the system.

\emph{Acknowledgments $-$}
We thank J. R. Glick, G.~Shchedrin, M. L. Wall, and L. Santos for useful discussions.
This work was performed with partial support of the NSF under grants
PHY-1806372, PHY- 1748958, OAC-1740130, CCF-1839232, and the AFOSR
under grant FA9550-14-1-0287. We acknowledge support of the UK Engineering
and Physical Sciences Research Council (EPSRC) through the `Quantum Science
with Ultracold Molecules' Programme (Grant No. EP/P01058X/1).
The authors acknowledge Colorado School of Mines supercomputing resources (Ref: https://ciarc.mines.edu/hpc/) made available for conducting the research reported in this paper.

\bibliography{refs}

\end{document}